\documentclass[useAMS,usenatbib]{mn2e}

\usepackage{graphicx,epsfig,subfigure}
\usepackage{url}

\usepackage{amsmath}
\usepackage{amssymb}

\usepackage{threeparttable}
\usepackage{float}
\usepackage{morefloats}

\title[On the dynamical evolution of the Orion Trapezium]
   {On the dynamical evolution of the Orion Trapezium}

    \author[Allen et al.]
  {Christine. Allen,$^1$\thanks{E-mail: chris@astro.unam.mx}
  Rafael Costero,$^1$ Alex Ruelas-Mayorga,$^1$ L. J. S\'anchez,$^1$
  \\
    $^1$Instituto de Astronom\'{\i}a, Universidad Nacional Aut\'onoma
     de M\'exico, Cd. Universitaria, M\'exico, D.F. 04510, M\'exico\\}

\begin{document}

\date{Accepted 2015. Received 2015 October; in original form 2015 October}

\pagerange{\pageref{firstpage}--\pageref{lastpage}}
\pubyear{2012}

\maketitle

\label{firstpage}


\begin{abstract}
 We discuss recent observational data on the transverse and radial velocities, as well as on the masses of the main components of the Orion Trapezium. Based on the most reliable values of these quantities we study the dynamical evolution of ensembles of multiple systems mimicking the Orion Trapezium. To this end we conduct numerical $N-$body integrations using the observed masses, planar positions and velocities, radial velocities, and random line-of-sight ($z$) positions for all components. We include perturbations in these quantities compatible with the observational errors. We find the dynamical lifetimes of such systems to be quite short, of the order of 10 to 50 thousand years. The end result of the simulations is usually a tight binary, or sometimes a hierarchical triple. The properties of the evolved systems are studied at different values of the crossing times. The frequency distributions of the major semiaxes and eccentricities of the resulting binaries are discussed and compared with observations.

\end{abstract}




\section{Introduction} \label{sec:intro}

The kinematics and dynamics of stellar systems composed of a few massive stars situated at comparable distances from each other have been little studied.  The prototype of such systems is, of course, the Orion Trapezium, and it is surprising that there are still many observational uncertainties associated with this famous and otherwise well studied system.  So, for instance, very few studies on dynamical stability of trapezium-type systems are found in the literature (Allen \& Poveda 1974a, 1974b, Pflamm-Altenburg \& Kroupa 2006). Allen \& Poveda (1974a) reported the results of numerical integrations of 30 systems, with trapezium-like initial positions and virialized velocities.  The main conclusion was that after 30 crossing times (about 1 million years) only  30\% of the systems survived as trapezia.  Allen \& Poveda (1974b) found that the probability of survival of a trapezium after 5,000 years was 14\%.  After $10^6$ years only about 4\% of the trapezia survived as such.  Pflamm-Altenburg \& Kroupa (2006) concluded that the Orion Trapezium was likely to be an OB star core in its final stages of decay.  Note that the results of those two studies are not directly comparable, since both used different criteria to define trapezium-type systems and their dynamical stability, as well as different initial configurations, and different values for the stellar masses. Neither attempted to use actual observed values of positions, velocities and masses of the Orion Trapezium components as initial conditions, due to the unavailability of such data.

It is clear that the region near the Orion Trapezium has been very active dynamically. Studies of the radio sources embedded in the BN-KL region (Rodr\'iguez et al. 2005, G\'omez et al. 2008) showed that three of these radio sources move away from a common point where they converged about 500 years ago.  Costero et al. (2008), observing only the radial velocity, found that Component E of the Orion Trapezium is escaping from it, probably as a result of dynamical interactions within the system.

The Orion Trapezium is composed of four very bright stars in a trapezoidal configuration,  and several fainter ones. The brightest star ($\theta^1$  Ori C) is responsible for the excitation of the Orion Nebula. It has an apparent magnitude $m=5.1$. $\theta^1$  Ori A, with a magnitude $m=6.7$, is an eclipsing binary. $\theta^1$  Ori B, with a magnitude $m=8$, is also known as BM Ori, and it is really a mini-cluster. $\theta^1$  Ori D has a magnitude $m=6.7$. The spectral types of the four brighter stars range from O to early B. $\theta^1$  Ori E is an escaping spectroscopic binary. Another star, $\theta^1$  Ori F has some times been considered as belonging to the Trapezium, but it is probably a foreground star. 

Observationally, the Orion Trapezium still presents many challenges. So, for instance, the radial velocities of its main components are  poorly known, probably as a consequence of the prevalence of multiplicity among them. Specifically, Component A is known to be an interferometric binary, A1+A3, while A1 itself is a spectroscopic binary with a separation of about 1 AU.  Hence, A is a triple system (Bossi et al. 1989, Close et al. 2013,). Component B is really a mini-cluster, composed by at least 5 stars in close proximity (Close et al. 2013).  The brightest component, star C is known to be at least a double, and probably attended by a third companion (Lehmann et al. 2010). Component D is a spectroscopic binary (Plaskett \& Pearce 1931). As mentioned before, Component E is a spectroscopic binary, with a systemic velocity sufficiently large to be escaping from the Trapezium (Costero et al. 2008).

With regard to the tangential motions, a recent study (Close et al. 2013) described in detail the internal motions of the mini-cluster $\theta^1$  Ori B.  Using these data and the best available radial velocities, Allen et al. (2015) conducted a numerical exploration of the dynamical evolution of this mini-cluster, arriving at the conclusion that its age is probably less than 30,000 years.

Recently, fairly reliable tangential motions for $\theta^1$  Ori itself have become available (Olivares et al. 2013). They were obtained by a combination of historical data and new measurements based on Hubble Space Telescope images.  With these new data, it seems worthwhile to conduct a  numerical exploration of the dynamical evolution of ensembles of systems resembling the best currently available observational data of the Orion Trapezium itself, similar to that we performed for the minicluster or sub-trapezium $\theta^1$ Ori B.

We discuss in Section 2 the data for the transverse velocities and their uncertainties. Section 3 is devoted to an examination of the available radial velocities and of the masses of the main Trapezium members. Section 4 describes the initial conditions and the numerical integrations performed to model the dynamical evolution of ensembles of Orion Trapezium-like systems.  The results are presented and discussed in Section 5. Section 6 summarizes our conclusions.

\section{Transverse velocities of the main Components of the Trapezium}
\label{sec:transversevelocities}

Olivares et al. (2013) obtained the relative motions of Components A to F of the Orion Trapezium (OT) using the diffracto-astrometry technique. They measured the evolution in time of the relative positions of the OT components. This was achieved by analyzing 44 high quality public-archive images of the Hubble Space Telescope Wide Field Planetary Camera 2. The images were taken over a span of 12 years (1995--2007). Due to the fact that the OT is the trapezium-type system best studied in the astronomical literature, they also used the historical compilation of OT data contained in the Washington Double Star Catalog (WDS) maintained by Mason et al. (2001), thus extending their analysis time-base to about 200 years.

For every pair of components they found the relative rate of separation as well as the temporal rate of change of their position angles. The relative rate of separation is expressed on the axes of an orthogonal coordinate system, where the positive $x$ direction corresponds to the west direction, while the positive $y$ direction corresponds to the north. In Table~\ref{tab:OTposrelC} we show the $x$ and $y$ positions  of the main Components A to E in a reference frame centered on the C component. Note that the errors of the respective positions are of the order of milliarcseconds.

\begin{table*}
\caption{Positions and velocities with respect to Component C.}
\begin{center}

\begin{tabular}{ccccc}

\hline

Star      & $x$   (+ West)            &       $y$  (+ North)              &           $v_{x}$ (+ West) )                  &	          $v_{y}$ (+ North)                                 \\
          &     (arcsec)              &           (arcsec)                &            (km\ s\textsuperscript{-1})        &	                            (km\ s\textsuperscript{-1})      \\

\hline
&   &   &   &   \\

A	      &  + 9.62                  &        +8.62                       &            $ -1.71 \pm 0.52$ 	              &                 $-1.54 \pm 0.47$           \\

B	      &  + 4.99                  &       +16.17                       &             $+0.41 \pm 0.26$                  &             	$+1.34 \pm 0.86$           \\

C         &    0.0                   &         0.0                        &              0.0                              &                  0.0                        \\

D         &  $-11.87$                &        +6.36                       &         	$-0.79 \pm 0.53$                  &             	$+0.43 \pm 0.28 $          \\

E	      &  +10.40                  &       +13.06                       &            $ +3.55 \pm 0.50 $                 &             	$+4.46 \pm 0.63 $         \\

\hline
\end{tabular}
\label{tab:OTposrelC}
\end{center}
\end{table*}

From Olivares et al. (2013) we take the values for the separation velocity between the components of the Orion Trapezium. A simple trigonometric projection of the separation velocities on the $x$ and $y$ axes is performed to obtain the relative velocities of the stars with respect to the C Component. These velocities are also shown in Table \ref{tab:OTposrelC}.

Since the precision of the historical measurements of the position angle (PA) is low, as stated in Olivares et al. (2013), the rates of change derived from the fits to the PA data are significantly less reliable than those obtained for the fits to the separation data. So, as was done in Olivares et al., we disregard in this paper the fitted PA rates of change. Note that this means that we are underestimating the transversal velocities.

Table~\ref{tab:OTposrelC} also shows the projected velocities in km\ s\textsuperscript{-1} in the $x$ and $y$ directions $(v_x, v_y)$. To obtain these values we adopted a distance to the Orion Nebula of $414 \pm 7$ pc (Menten et al. 2007).

We calculated the temporal rate of change of separation velocity and projection angle with respect to the C component as explained above. The results are summarized in Table~\ref{tab:ROCS}. We should mention that we became aware of a typing error in the sign of the separation velocity of the E component in Table 4 of the above cited article: instead of -5.7, the correct value is +5.7 km\ s\textsuperscript{-1} (Olivares et al. 2016).

\begin{table}
\caption{Temporal rate of change of separation velocity and projection angle.}
\begin{center}

\begin{tabular}{ccc}
\hline
Pair              & Separation Velocity    & Projection Angle    \\
          &  (km\ s\textsuperscript{-1} )    & (degrees)    \\
\hline

CA  &$   -2.3  \pm   0.7  $  &   $     221.9 ^\circ   $ \\
CB  &$   +1.4  \pm   0.9  $  &   $      72.9^\circ    $ \\
CD  &$   +0.9  \pm   0.6  $  &   $    151.8^\circ    $ \\
CE  &$   +5.7  \pm   0.8  $  &   $     51.5^\circ    $ \\

\hline
\end{tabular}
\label{tab:ROCS}
\end{center}
\end{table}

In Table~\ref{tab:ROCS}, the first column indicates the pair of OT components considered, the second column lists the rate of change in separation (in km\ s\textsuperscript{-1}) and the third column  the projection angle of the separation velocity vector measured clockwise with respect to the west direction on the sky. The error in the separation velocity was estimated using the projected $v_x$ and $v_y$ errors added in quadrature.

As a check of our calculations it is worth mentioning the radio determination of the relative motion of Component E with respect to Component A made by Rodr\'iguez (2008). Using archival VLA data he was able to measure the relative displacement of Components A3 (the interferometric companion to A) and E.  With an adequate correction for the motion of A3 around A, a total separation velocity for AE of $9.1 \pm 1.0$ km\ s\textsuperscript{-1} was found, in approximately the NW direction. Our data produce a separation velocity for AE of $8.0 \pm 1.0$ km\ s\textsuperscript{-1}, also in approximately the NW direction, and thus agree, within the uncertainties, with the radio determination.   Finally, we note that the results presented in this section are fairly consistent with those published by Allen et al. (2004), based solely on historical measures.

\begin{table}
\caption{Radial velocities of Trapezium stars with respect to Component C. The radial velocity of C was taken to be $27.3 \pm 1$ km\ s\textsuperscript{-1}.}
\begin{center}
\begin{tabular}{cc}
\hline
Name & $v_r$\\
     &(km s\textsuperscript{-1})\\
\hline
$\theta^1$ Ori A & $0.7 \pm 1.0$\\
$\theta^1$ Ori B & $-1.3 \pm 1.0$ \\
$\theta^1$ Ori C & $0 $ \\
$\theta^1$ Ori D & $5.1 \pm 3.0$\\
$\theta^1$ Ori E & $7 \pm 1.0 $ \\
\hline
\end{tabular}
\end{center}
\label{tab:rvt}
\end{table}

\section{The Radial Velocities and Masses of the Bright Orion Trapezium Members}
\label{sec:massesandradialvels}

The masses and radial velocities of the Trapezium members are difficult to measure because almost all of them are multiple systems, immersed in bright nebulosity. In addition, their mutual proximity complicates the observation of the weakest components.

\subsection{The Radial Velocities}
\label{subsec:radialvelocities}

The brightest, hottest and most massive star in the Orion Trapezium, $\theta^1$ Ori, is  Component C = HD 37022.  It was suspected to be a spectroscopic binary since the first extensive spectroscopic surveys by Plaskett \& Pearce (1931) and Frost, Barrett \& Struve (1926).  More recently it was found to be an oblique magnetic rotator, a property that causes noticeable periodic variations of the equivalent widths and profiles of certain strong spectral lines (Stahl et al. 1993, 1996) with a period equal to that of the star's rotation (Sim\'on-D\'iaz et al. 2006). Attempts to relate the observed additional radial velocity variability to orbital motions were made by Vitrichenko (2002) and Stahl et al. (2008), after the star was identified to be an interferometric binary by Weigelt et al. (1999). Now, after the first complete 11-year orbit has been covered with both interferometric and radial velocity observations (Balega et al. 2014, 2015), precise values for the mass and systemic radial velocity of the binary should be available.  However, even with sufficient radial velocity data for the primary component (Stahl et al. 2008, Lehmann et al. 2010, Grellmann et al. 2013) and  for the secondary (Balega et al. 2014, 2015), the combined astrometric and spectroscopic orbital solution to the data is not satisfactory, as admitted by the latter authors.  Indeed, uncertainties in the observational data, especially in the radial velocities of both interferometric components --which could be due to non-photospheric contributions (Stahl et al., 2008) or to additional stellar light sources in the primary (Vitrichenko et al. 2011)-- as well as from the large projected rotation velocity of the secondary component (Balega et al. 2014, 2015), introduce a large dispersion in the observed velocity curve and, hence, in the orbital parameters.

The systemic radial velocity ($\gamma$) of $\theta^1$ Ori C has been determined by Kraus et al. (2009) to be $23.6$ km\ s\textsuperscript{-1}, by adjusting the radial velocities observed by Stahl et al. (2008) for the primary component to those predicted by their computed interferometric orbit. Lehmann et al. (2010) obtained $\gamma = 25 \pm 4$ km\ s\textsuperscript{-1} after correcting for the modulations observed in the radial velocity of the primary component caused by the oblique magnetic rotator (semi amplitude $\sim$ 6 km\ s\textsuperscript{-1}, P $\sim$ 15.3 d) and by a putative closer component (semi amplitude $\sim$ 17 km\ s\textsuperscript{-1}), P $\sim$ 61 d). In their recent papers, Balega et al. (2014, 2015) calculated $\gamma = 31.0\pm2.0$ km\ s\textsuperscript{-1} and $\gamma = 29.4\pm0.6$ km\ s\textsuperscript{-1}, based on their combined interferometric and spectroscopic orbital solution which included the radial velocities of both the primary and the secondary interferometric components near periastron. These authors used spectra with an extremely large signal-to-noise ratio that allowed the radial velocity of the secondary component to be measured. No comment on the possible existence of an additional component is made in these papers.

We have adopted $V_{rad} = 27.3$ km\ s\textsuperscript{-1} for the systemic radial velocity of Component C. This value is representative of all the above mentioned velocities and equal to the average of smallest and the largest (23.6 and 31 km\ s\textsuperscript{-1}). Furthermore, if either the smallest (or the largest) value were used,  Components D (or B) would have a radial velocity large enough to escape from the stellar group.

 Some years ago, $\theta^1$ Ori A = V1016 Ori = HD 37020 was discovered to be an eclipsing and spectroscopic binary (Lohsen 1975, 1976; Bossi et al. 1989) in a very eccentric, $P = 65.45$ d orbit. Its systemic velocity was determined by Vitrichenko, Klochkova, \& Plachinda (1998) and by Stickland \& Lloyd (2000); both groups used very similar archival data and a few additional measurements of their own so, not surprisingly, the two research teams reach nearly equal orbital parameters, including the systemic radial velocity, $28$ km\ s\textsuperscript{-1}.  We adopted this value for $\theta^1$ Ori A.

The orbital parameters of $\theta^1$ Ori B = BM Ori = HD 37021, which is itself an eclipsing and spectroscopic binary in a nearly circular, $P=6.45$ d orbit,  have been derived by several authors.  However, they differ significantly from each other, especially for the value of the semi amplitude of the velocity curve of the primary component and that of the systemic velocity. Using more precise data, Vitrichenko \& Klochkova (2004) present convincing evidence in favor of their interpretation of these differences as due to a third component in a long period, highly eccentric orbit. We adopt the systemic velocity obtained by these authors for the putative hierarchical triple system, namely $26$ km\ s\textsuperscript{-1}. This value is somewhat larger than that obtained from averaging the previously published systemic velocities of the eclipsing binary ($\sim 24$ km\ s\textsuperscript{-1}),  but probably more exact and certainly closer to that of the other Trapezium members. The systemic velocity of the minicluster will be assumed to be that of BM Ori.

$\theta^1$ Ori D = HD 37023 has been known to be a spectroscopic binary ever since the Plaskett \& Pearce (1931) first noticed large radial velocity variations in this star. However, to our knowledge, only Vitrichenko (2000) has gathered and selected published measures of the radial velocity and, together with few more data from archival IUE spectra of this star, he attempted to find the periodicity of the observed variations. That work resulted in two probable orbital periods, differing by a factor of about two; Vitrichenko prefers the shorter period (about 20.3 d), from which he obtains a systemic velocity of $\gamma = 32.4\pm1.0$ km\ s\textsuperscript{-1}.  This is the value we have adopted for the radial velocity of this object. The error given in Table~\ref{tab:rvt} is representative of the difference in the solutions for $\gamma$ from the two possible periods.

$\theta^1$ Ori E was discovered to be a double lined spectroscopic binary by Costero et al. (2006). Its systemic velocity was obtained by Herbig \& Griffin (2006) and, using many more data points (80 vs 10), by Costero et al. (2008). The former authors obtained $30.4 \pm 1$ km\ s\textsuperscript{-1} and the latter $34.3 \pm 0.7$ km\ s\textsuperscript{-1}. We have adopted this last value.

In Table~\ref{tab:rvt} we list the values of the radial velocities for stars A to E used in this paper as well as their corresponding adopted errors.

\subsection{The Masses}
\label{subsec:masses}

\subsubsection{The Mass of the most massive component}

\begin{table}
\caption{Masses of Trapezium stars.}
\begin{center}
\begin{tabular}{cc}
\hline
Name & Mass (M$_\odot$) \\
\hline
$\theta^1$ Ori A & $27  \pm 1.35$  \\
$\theta^1$ Ori B & $15 \pm 0.75$  \\
$\theta^1$ Ori C & $45 \pm 10$ ; \space  $65 \pm 3.25$  \\
$\theta^1$ Ori D & $25 \pm 1.25$ \\
$\theta^1$ Ori E & $7 \pm 0.35 $  \\
\hline
\end{tabular}
\end{center}
\label{tab:mt}
\end{table}

$\theta^1$ Ori C is the only Trapezium member for which the orbital inclination is known. In a recent paper, Balega et al. (2015) obtained the total mass of the interferometric binary trough the simultaneous spectroscopic and interferometric orbital solution; they found $M_C = M_{C1} + M_{C2} = 45.5 \pm 10.0 M_\odot$. The estimated mass ratio, $q = M_{C2}/M_{C1} = 0.36 \pm 0.05$, implies $M_{C1} \simeq 33 M_\odot$ and $M_{C2} \simeq 12M_\odot$. In spite of the uncertainties, these values are in good agreement with those predicted by the models by Martins et al. (2005) for a single O5-7 main-sequence star, with an effective temperature of $39,000\pm1,000 K$ (as measured by Sim\'on-D\'iaz et al. 2006 for the primary component)  and a B0V secondary with an effective temperature of 30,000 K  (as estimated for the interferometric companion by Balega et al. 2015). Those models take into account the effects of non-LTE, stellar wind, and line-blanketing  and agree with recent calibrations of stellar parameters for different spectral types by Torres et al. (2010) and Nieva \& Przybilla (2014).

Another close component, of smaller mass, has been suggested by Vitrichenko (2002) for $\theta^1$ Ori C. After filtering for the effects of the oblique magnetic rotation, Stahl et al. (2008) obtained an orbital period $\simeq 61.5$ d (four times the rotation period of the massive star) for this component. This result was confirmed by Lehmann et al. (2010), who additionally estimated its mass to be larger than $1 M_\odot$.  If indeed real, this putative component would be included in the dynamical mass of the interferometric binary (Balega et al. 2015) but, of course, not in the spectroscopic mass of the primary (Sim\'on-D\'iaz et al. 2006).

An intermediate mass for the interferometric binary is derived from Sim\'on-D\'iaz et al. (2006) determination of the spectroscopic mass for the brightest component, $M_{C1} = 45 M_\odot$. Since the secondary component is a B0V star according to Balega et al. (2014, 2015), its mass should be $M_{C2} \simeq 20 M_\odot$ if the generally accepted calibration of spectral type with physical parameters of massive stars by Vacca, Garmany \& Shull (1996) is adopted. This adds up to a total mass of about $65 M_\odot$ for Component C.

It should be noted, however, that Vitrichenko et al. (2012) obtained $6.8 \pm 0.4 M_\odot$ for the mass of this component, from the ratio of the semi-amplitudes of the velocity curves of the primary and the putative component, and assuming the mass of the the primary to be $48\pm0.4 M_\odot$. This would imply a total mass of about $75 M_\odot$ for the Trapezium C system.

For these reasons, we initially adopted 45 $M_\odot$ for the total mass of $\theta^1$ Ori C (considered as a single mass point) for the dynamical simulations described in Section \ref{sec:dynamicalmodel}. However, as we will discuss in Section \ref{sec:results}, this value for the mass resulted in the rapid destruction of the Trapezium in less than a crossing time (9,454 years).  Therefore we also considered larger masses for the C system, compatible with older determinations but still within the margin allowed at present by the observational uncertainties of this difficult system.

In summary, besides adopting $M_C = 45 M_\odot$ for the total mass of $\theta^1$ Ori C (Balega et al. 2015), we also performed numerical integrations using the larger value of $M_C = 65 M_\odot$ (Sim\'on-D\'iaz et al. 2006) for the combined binary mass. We also ran simulations using $M_C = 70 M_\odot$. The results of this third ensemble were very similar to the ones with $M_C = 65 M_\odot$ and will not be further discussed here.

\subsubsection{The masses of the other components}

The masses of both components of the eclipsing binary $\theta^1$ Ori A were obtained by Vitrichenko \& Plachinda (2001), based on previously calculated orbital elements derived from the velocity curve of the primary component (Vitrichenko, Klochkova \& Plachinda, 1998) and on a single radial velocity measure of the secondary component obtained during an eclipse.  The former authors find $M_{A1} = 21 \pm 4 M_\odot$ and $M_{A2} = 3.9 \pm 0.4 M_\odot$  for the masses of the primary and secondary components. The value for the primary is somewhat large compared with the spectral or evolutionary mass (about 15 $M_\odot$), as  given by Sim\'on-D\'iaz et al. (2006) or Nieva \& Przibilla (2014). Additionally, there is a suspected third, distant, component interferometrically identified by Petr et al. (1998) about 0.2 arcsec north of the eclipsing binary. This component was also associated with the strong, non-thermal radio source first detected by Churchwell et al. (1987). Close scrutiny of its relative transverse motion provides strong evidence in favor of this star being  gravitationally linked to the eclipsing binary, though it is not yet possible to discard its being a chance intruder. From its IR photometric properties and assuming it belongs to the Trapezium Cluster, the mass of this third component was estimated to be about 4 $M_\odot$ by Weigelt et al (1999), Shertl (2003), Grellmann et al. (2013). Furthermore, Close et al. (2012) estimate $M_{A1} = 20 M_\odot$ and $M_{A2} = 2.6 M_\odot$ and $M_{A3} = 4 M_\odot$. For these reasons, we adopted for the dynamical modelling a total mass of 27 $M_\odot$ for the combined system $\theta^1$ Ori A.

The mini-cluster $\theta^1$ Ori B consists of a spectroscopic binary (or even a triple system),  with three addi\-tio\-nal interferometric companions.  For these components we adopted in our previous paper (Allen, Costero and Hern\'andez, 2015), the approximate masses given by Close et al. (2012), namely:  3$M_\odot$, 2.5$M_\odot$ and 0.2 $M_\odot$, for Components 2, 3 and 4, respectively, using these authors' nomenclature. However, for the present numerical simulations we prefer to use the masses of the primary and secondary components of the eclipsing binary and the putative third spectroscopic companion given by Vitrichenko, Klochkova \& Tsymbal (2006), namely 6.3$M_\odot$, 2,5$M_\odot$ and 1.8$M_\odot$. The mass given for the se\-con\-da\-ry component (B5) of the eclipsing binary  by Close et al. (2012) is erroneous;  we used it in our previous paper with due reservations.  The total mass there adopted was 19.7 $M_\odot$. With the present considerations in mind, the total mass of the $\theta^1$ Ori B mini-cluster would add up to about 16.3$M_\odot$.  Since the putative third spectroscopic component has not been independently confirmed, we adopted here a total mass of 15$M_\odot$ for this quintuple (or sextuple) system.

Concerning the mass of the $\theta^1$ Ori D binary system we adopted a mass of $M = 25M_\odot$ for the  for the following reasons: (1) The spectroscopic and evolutionary mass of the primary component was estimated to be about 18$M_\odot$ by Sim\'on-D\'iaz et al. (2006) and Voss et al. (2010).   The former authors estimate the temperature of this component to be 32,000 K.  (2) One of us (RC) has obtained several high S/N Echelle spectra of this star, a few of which show a doubling of the Si III triplet lines around 4,560 \AA (probably when the binary is near quadrature); the equivalent width of each weak component is roughly 1/4 that of its corresponding bright one, and is clearly blue-shifted with respect to the primary. The existence of these faint Si III components imply that the temperature of the secondary star must be hotter than 15,000 K, since these ions rapidly vanish at lower temperatures. However, the  relative intensities of the two spectral systems require the temperature for the secondary component to be about 0.75 that of the primary (if both components are of about the same size);  that is,  the secondary should have a temperature of about 24,000 K,  high enough to also allow the observation of the doubling of strong O II lines, which become very weak at 20,000 K. This doubling is not observed, so the temperature of the secondary must be close to 20,000 K.  This corresponds to a B2 main-sequence star and, according to the calibrations by Torres et al. (2010) and Nieva \& Przybilla (2014), this would imply a mass around 7$M_\odot$. Consequently, the adopted mass for the $\theta^1$ Ori D system is the sum of the mass of its two components, obtained as described above, namely 25$M_\odot$.

The mass of the double-lined spectroscopic binary $\theta^1$ Ori E is the best known among the Trapezium members. Morales-Calder\'on et al (2011, 2012), as a part of a very complete {\it Spitzer} 3.5$\mu$m and 4.5$\mu$m photometrical survey designed to study short and long term variability, established that $\theta1$ Ori E is a grazing eclipsing binary (maximum depth of about 0.07 mag); using the observed light curve and the radial velocities by Costero et al. (2008), Morales-Calder\'on et al. (2012) obtained physical parameters of the almost identical components of the binary, including their individual masses ($2.80 \pm 0.05 M_\odot$), in good agreement with Herbig \& Griffin (2006) who, from the location of the components on pre main-sequence evolutionary tracks, estimated the mass of each component to be about 3.5$M_\odot$. We adopted 7$M_\odot$ for the total mass of the binary.

In Table~\ref{tab:mt} we list the values of the masses we adopted in this paper for stars A to E.

\section{The Dynamical Model}
\label{sec:dynamicalmodel}

To realistically model the dynamical evolution of the Orion Trapezium by N-body simulations we require as initial conditions the best available values for the positions, transverse velocities, radial velocities and masses of the main components.  The membership of Component F to the Orion Trapezium is doubtful (see Olivares et al. 2013 and references therein). It is probably a foreground star (Alves \& Bouy, 2012). Therefore, we considered only components A, B, C, D, and E.

For the dynamical model we took a Monte Carlo approach, perturbing each observed value by amounts compatible with the observational errors. We assumed that all components act as point masses, disregarding the subsystems. As discussed in Section \ref{sec:massesandradialvels}, we are indeed aware of the fact that most OT components are in fact binaries or multiples, but their separations are much smaller than the closest encounters during the integrations. We checked the closest encounters that actually occured during the integrations and found them to be at least an order of magnitude larger than the widest inner binary, so our assumption is justified. For the distance to the Orion Trapezium we adopted the value of Menten et al. (2007) for the Orion Nebula Cluster, that is, 414 pc.

The positions on the plane of the sky are accurately known. Hence, for the perturbations in the positions we assumed a Gaussian distribution with a dispersion of only 8 AU.  To assign  the $z$-positions we assume again a Gaussian distribution around zero, with a dispersion of 1,200 AU.   This value was chosen in order to preserve the trapezium configuration also in the $z$-direction. With such a dispersion the $z$ separations hardly ever exceed the planar separations between the components and all resulting systems turned out to be bound. They all resemble, in projection, the actual Orion Trapezium.

The transverse velocities were discussed in Section \ref{sec:transversevelocities}.  We take the values there given, with perturbations in the velocities compatible with the quoted uncertainties.  The perturbations were again assumed to have a Gaussian distribution with a dispersion equal to the uncertainty for each component.

As discussed in Section \ref{sec:massesandradialvels},  in order for the main components A, B, C, and D to form a bound system, we assumed a radial velocity of 27.3 km\ s\textsuperscript{-1} for Component C, midway between the determinations of $23.6$ km\ s\textsuperscript{-1} (no error quoted)(Kraus et al. 2009) and $31.0 \pm 2$ km\ s\textsuperscript{-1} (Balega et al. 2014). Both determinations are quite uncertain.  We note that using the values given by these authors we find that one or more stars (apart from Component E) escape from the system even before starting the integrations;  this situation appears to us to be implausible, and for this reason we adopted the mean between the two values for the radial velocity of C.  Each radial velocity received perturbations with a Gaussian distribution and a dispersion fixed by the observational errors.

As stated in Section \ref{sec:massesandradialvels}, the masses for the components are also very uncertain, especially that of Component C.   We will discuss below the influence of the uncertain value of the mass of Component C on the dynamical evolution of the simulated systems.
With the values and the perturbations just discussed, we generated initial conditions for three ensembles of 100 systems, each ensemble having a different value for the mass of Component C.
All generated sets of initial conditions resulted in bound systems, with slightly positive values for the virial ($2T+V$, where $T$ is the kinetic and $V$ the potential energy). In all cases, Component E had a positive energy, confirming its status as an escaping star (Costero et al. 2008).  Note that if we disregard star E we obtain values for the virial close to zero.
For the numerical N-body integrations we used the well-tested code by Mikkola \& Aarseth (1993).  This code implements a chain regularization algorithm, which allows to accurately follow the close encounters expected to occur in such few-body systems.  Typical values for the relative error in the energy, $\Delta E/E$, at the end of the integrations (after 100 crossing times) were of the order of $10^{-11}$, those for the angular momentum an order of magnitude smaller.  We obtained ``snapshots'' of the systems at 1, 5, 10, 25, 50, and 100 crossing times. At 100 crossing times (about 1 million years) the dynamical evolution is practically over.

\section{Results}
\label{sec:results}

\subsection{The Lifetimes of the Modeled Systems}
\label{subsec:lifetimes}

\begin{figure*}
\centering
\includegraphics[width=17.0cm,height=10.0cm]{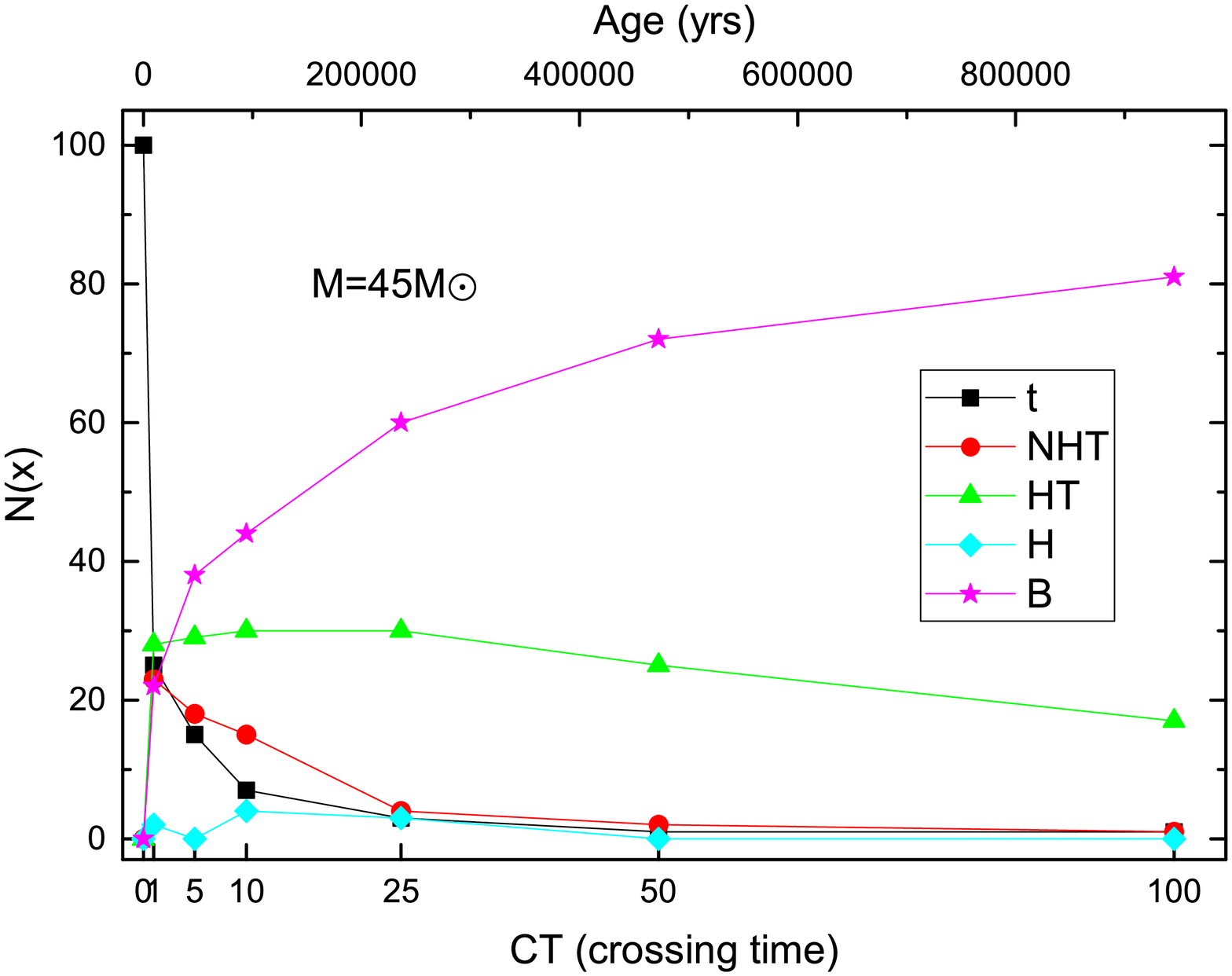}
\caption{Number of different systems generated from 100 initial trapezia ($M_C = 45M_\odot$) as a function of crossing time or age. t stands for trapezia of both types (1, 2, 3, 4) and (1-2, 3, 4), NHT for Non-Hierarchical Triples, HT for Hierarchical Triples, H for Hierarchical Quadruples and B for Binaries.}
\label{fig:TEORICOS_MC45}
\end{figure*}

We first conducted numerical integrations of an ensemble of 100 systems with initial conditions as described above and adopting $45 M_\odot$ for the mass of Component C (Ensemble 1).  In all cases, Component E quickly escaped, leaving only a system of four stars for the remaining dynamical evolution.  We discuss later the behavior of Component E.
To assess the persistence of trapezium systems as such, and to be able to estimate their mean lifetimes, we recall here the traditional definition of trapezia given by Ambartsumian (1954):  let a multiple systems (of 3 or more stars) have distances $ab$, $ac$, $bc$, etc, between its components.  If three or more such distances are of the same order of magnitude, then the multiple system is of trapezium type.  Otherwise, it is of hierarchical type.  In this context, two distances are ``of the same order of magnitude" if their ratio is greater than 1/3 but smaller than 3.
The results of our first experiment (Ensemble 1), which assumes a mass for Component C of $45 M_\odot$, after 1, 5, 10, 25, 50 and 100 crossing times are displayed in Table~\ref{tab:TEORICOS45}  (one crossing time corresponds to 9,454 years). The table clearly shows that already after about 10,000 years, only 5 systems survived as trapezia resembling the original configuration.  We denote such systems as trapezia of type (1, 2, 3, 4).  In another 20 cases a close double was formed, but the other two stars remained bound, with trapezium-like separations.  We call such systems trapezia of type (1-2, 3, 4).  Of course, they do not resemble the original form of the Orion Trapezium.
The results are even more dramatic for longer crossing times. After 100 crossing times (about one million years) 81 systems completely dissolved, leaving only binaries.  No trapezium of the original type (1, 2 ,3, 4) survived.  Among the end products we found only one trapezium of type (1-2, 3, 4), as well as one non-hierarchical triple and 17 hierarchical triples.  These results are shown in Table~\ref{tab:TEORICOS45} and displayed in Figure \ref{fig:TEORICOS_MC45}.

\begin{table*}
\begin{center}
\caption{Number of systems resulting at increasing crossing times for $M_C = 45M_\odot$.}
\begin{tabular}{ccccccc}
\hline

Crossing Times	&	Trapezia (t) 	&   Trapezia (t)   &	Non-Hierarchical	&	Hierarchical 	&	Hierarchical 	&	Binaries	\\
                &  (1, 2, 3, 4)     &   (1-2, 3, 4)    &     Triples (NHT)      &   Triples (HT)    &   Quadruples (H)  &      (B)      \\
\hline
                &                   &                  &                        &                   &                   &               \\
          0  	&	 100	        &        0         &	           0	    &	            0	&        	0	    &     	0	    \\
          1	    &	   5	        &       20         &	          23	    &	           28   &	        2	    &   	22   	\\
          5   	&	   6	        &        9         &	          18        &	           29	&	        0	    &	    38	    \\
         10	    &	   4	        &        3         &	          15	    &	           30   &        	4	    &	    44   	\\
         25	    &	   1	        &        2         &	           4	    &	           30	&	        3	    &	    60   	\\
         50	    &	   0	        &        1         &	           2	    &	           25	&	        0	    &	    72  	\\
        100	    &	   0         	&        1         &	           1        &	           17	&        	0	    &	    81   	\\
\hline

\end{tabular}
\label{tab:TEORICOS45}
\end{center}
\end{table*}

The results of this first experiment (Ensemble 1)  would appear to contradict the results obtained by Allen \& Poveda (1974) who found lifetimes of about 500,000 years for their systems.  Note, however, that they worked with larger values for the masses of the components.  Their systems were composed of two stars of $50 M_\odot$, two of $20 M_\odot$ and two of $15 M_\odot$.  On the other hand, these results agree (at least qualitatively), with those found by Pflamm-Altenburg \& Kroupa (2006) who concluded that the Orion Trapezium should have dissipated by now.

For a quantitative assessment, we define the ``lifetime" of a trapezium as the time necessary for one half of the original population to have evolved into non-trapezium configurations.  The outcome of the first experiment (Ensemble 1) implies a lifetime for systems resembling the Orion Trapezium of much less than 10,000 years.  Table \ref{tab:TEORICOS45} shows that, indeed, after one crossing time only 25 trapezium-like systems survived, out of which only 5 resembled the original configuration.  We recall that the lifetime of an O star is about a million years.  Compared to this lifetime, our present systems, which are designed to resemble the actual Orion Trapezium,  dissolve in extremely short times.  This would imply an extreme youth for Orion Trapezium-like systems.  Since we consider such a possibility unlikely, we conducted experiments with different values for the mass of Component C.

\begin{figure*}
\centering
\includegraphics[width=20.0cm,height=10.0cm]{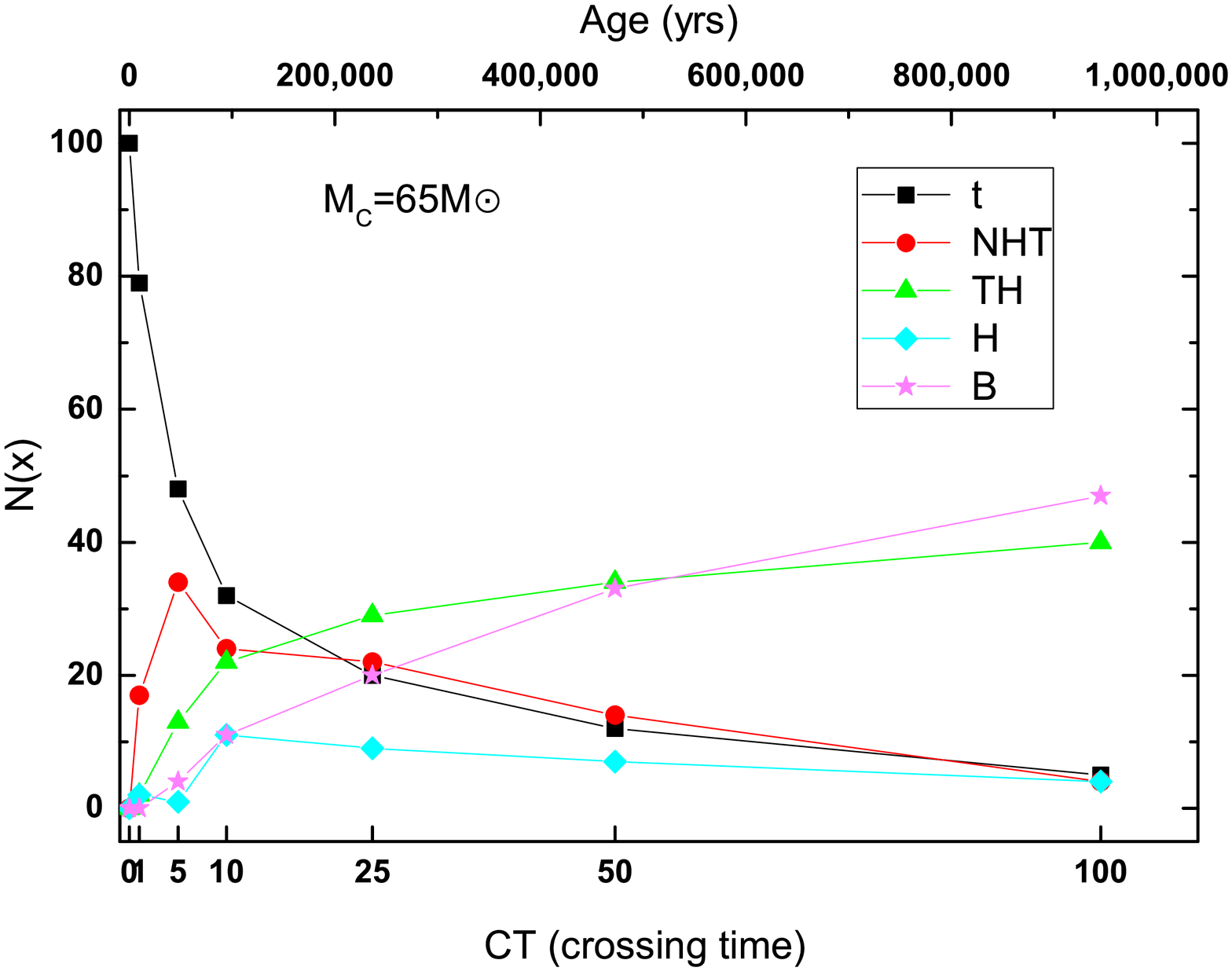}
\caption{Number of different systems generated from 100 initial trapezia ($M_C = 65M_\odot$) as a function of crossing time or age. Labels as in Fig. \ref{fig:TEORICOS_MC45}.}
\label{fig:TEORICOS_MC65}
\end{figure*}

The second experiment (Ensemble 2) consisted of a further 100 systems with the same initial conditions as above, but with a mass of $65 M_\odot$ for Component C.  The results of this experiment are shown in Table \ref{tab:TEORICOS65}, after 5, 10, 15, 25, 50 and 100 crossing times (corresponding to about 10,000,  50,000, 100,000, 250,000, 500,000, and one million years, respectively), and depicted in Figure \ref{fig:TEORICOS_MC65}. Now we can see that after about 10,000 years 41 systems survive as trapezia of type (1, 2, 3, 4), while a further 38 become trapezia (1-2, 3, 4).  Hence, the dynamical lifetimes in this case are clearly greater than 10,000 years.  The table also shows that after 5 crossing times 22 systems survive as trapezia (1 ,2, 3, 4), while another 26 became trapezia (1-2, 3, 4). Hence, the dynamical lifetimes are between 10,000 and 50,000 years, a more plausible outcome.  We recall that the dynamical age estimated for the minicluster $\theta^1$ Ori B was found to be about 30,000 years (Allen et al. 2015). Thus, these results on the dynamical age of systems resembling the Orion Trapezium are compatible with the age we found for one of its components.

The third experiment (Ensemble 3) consisted of another 100 systems, but now taking a more extreme value, $M_C = 70 M_\odot$. The results were very similar to those obtained with $M_C = 65 M_\odot$ and will not be further discussed.

\begin{table*}
\begin{center}
\caption{Number of systems resulting at increasing crossing times for $M_C = 65M_\odot$.}
\begin{tabular}{ccccccc}
\hline

Crossing Times	&	Trapezia (t) 	&   Trapezia (t)   &	Non-Hierarchical	&	Hierarchical 	&	Hierarchical 	            &	Binaries	\\
                &  (1, 2, 3, 4)     &   (1-2, 3, 4)    &     Triples (NHT)      &   Triples (HT)    &   Quadruples (H)              &      (B)      \\
\hline
                &                   &                  &                        &                   &                               &               \\
          0  	&	 100	        &	     0         &        0	            &	    0	        &        	0	                &     	0	    \\
          1	    &	  41	        &	    38         &       17	            &	    2	        &	        2	                &   	0   	\\
          5   	&	  22	        &	    26         &       34            	&	   13	        &	        1	                &	    4	    \\
         10	    &	  12	        &	    20         &       24	            &	   22	        &          11	                &	    11   	\\
         25	    &	   6	        &	    14         &       22	            &	   29	        &	        9	                &	    20   	\\
         50	    &	   2	        &	    10         &       14	            &	   34	        &	        7	                &	    33  	\\
        100	    &	   0         	&	     5         &        4               &	   40	        &        	4	                &	    47   	\\
\hline

\end{tabular}
\label{tab:TEORICOS65}
\end{center}
\end{table*}

It is interesting to ask the question of whether the escaping Component E could plausibly have been  bound to the Trapezium in the past. To test this possibility, starting from the same initial conditions we ran the integrations backward in time. We found that in 25\% of the cases, Star E acquired a negative energy (that is, became bound) at times varying between 0.1 and 0.3 crossing times (about 1,000 to 3,000 years).  We can conclude that Star E probably was bound to the Orion Trapezium in the recent past, and that it escaped only about 2,000 years ago.

\begin{figure*}
\centering
\includegraphics[width=20.0cm,height=10.0cm]{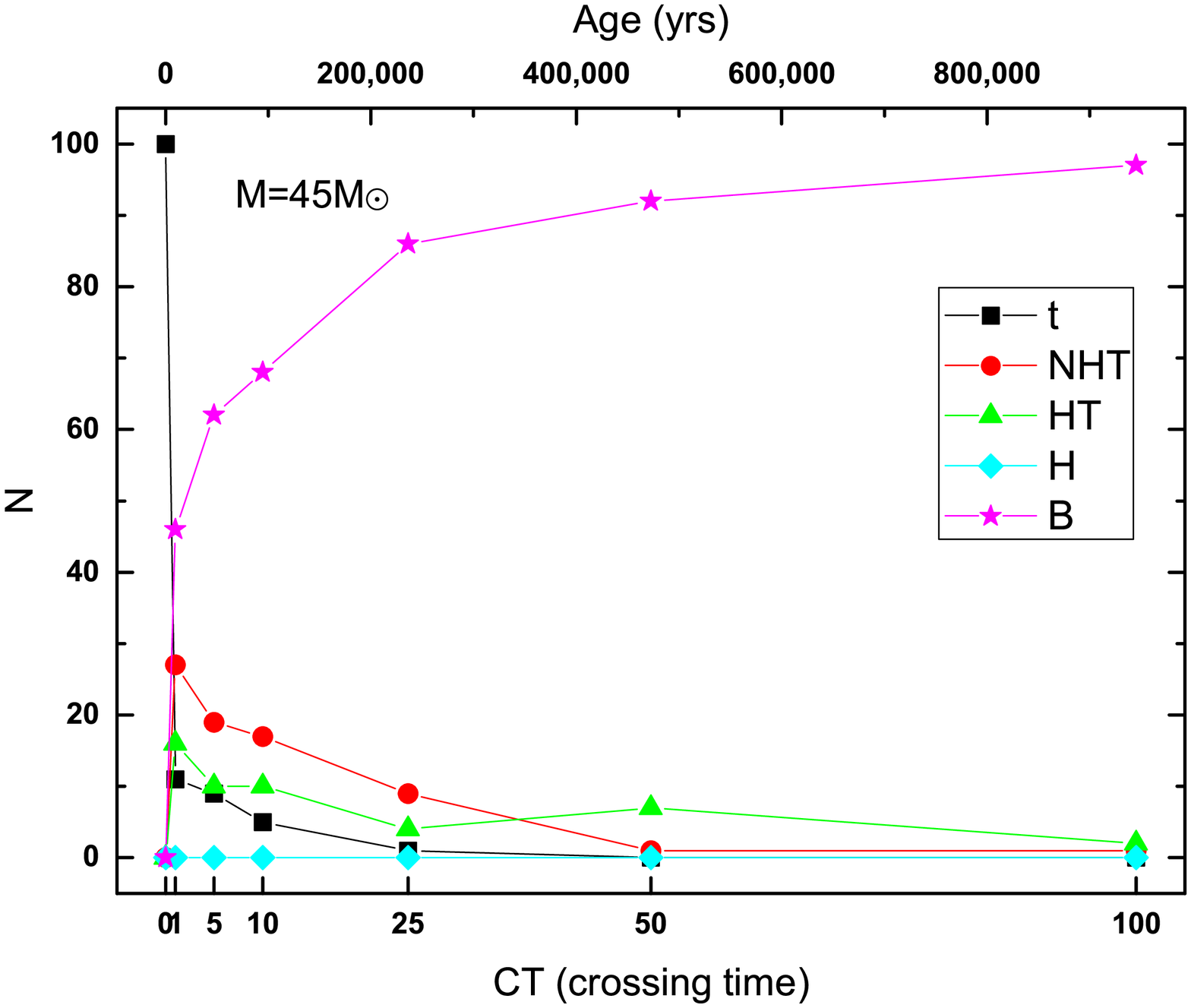}
\caption{Number of different ``observable" systems generated from 100 initial trapezia ($M_C = 45M_\odot$) as a function of crossing time or age. Labels as in Fig. \ref{fig:TEORICOS_MC45}.}
\label{fig:OBSERVABLES45}
\end{figure*}

\begin{figure*}
\centering
\includegraphics[width=20.0cm,height=10.0cm]{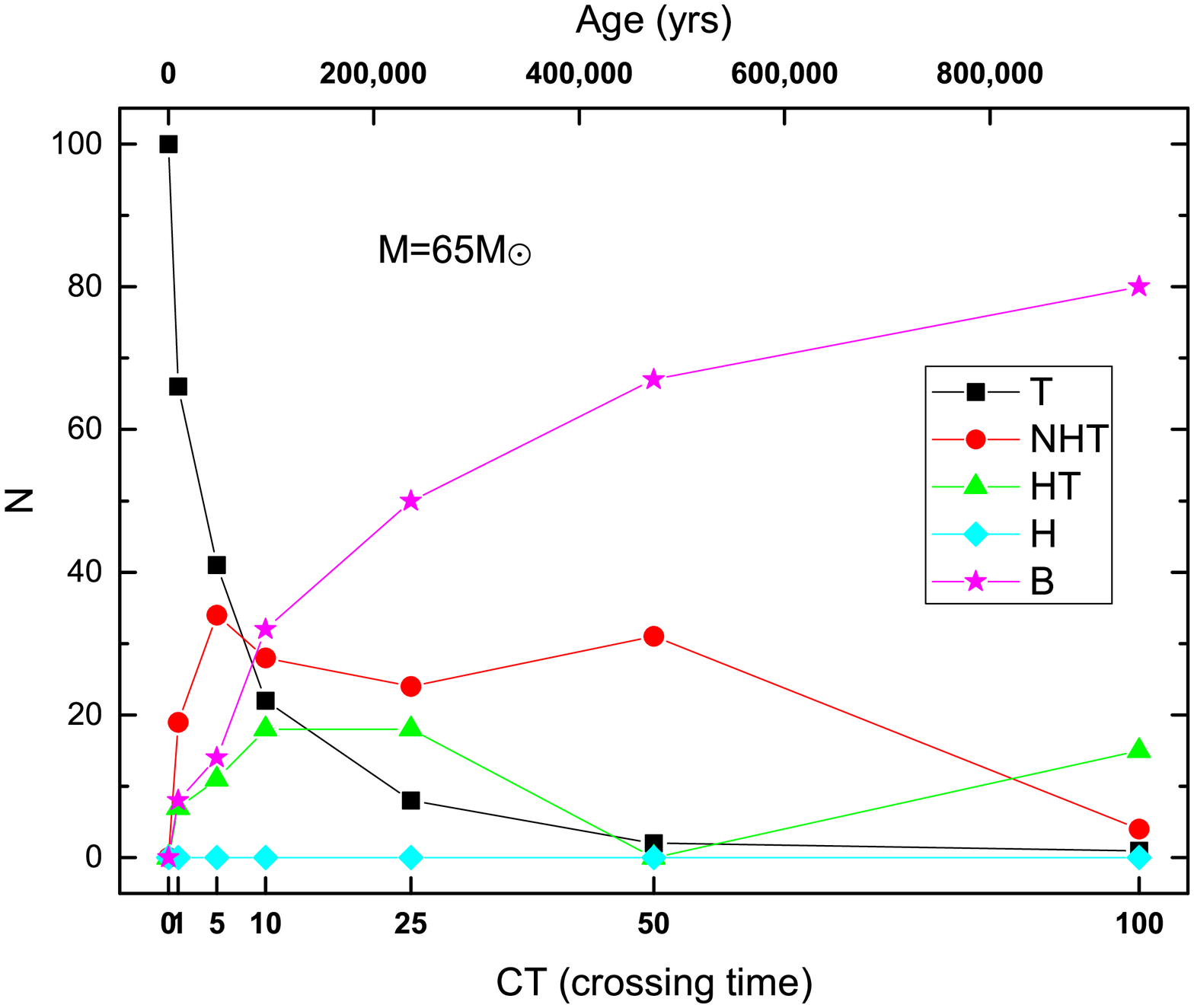}
\caption{Number of different ``observable" systems generated from 100 initial trapezia ($M_C = 65M_\odot$) as a function of crossing time or age. Labels as in Fig. \ref{fig:TEORICOS_MC45}.}
\label{fig:OBSERVABLES65}
\end{figure*}

We now discuss how the systems resulting from the numerical simulations would look to observers.  For this purpose, we assume an upper limit  of 40,000 AU for the separation of a component to still be detectable as a member of the system.  At the adopted distance for the Orion Trapezium this limit corresponds to an angular separation of about 97 arcsec.  If we now disregard all components with separations larger than this limit, we obtain the results displayed in Tables \ref{tab:OBSERVABLES45} and \ref{tab:OBSERVABLES65}, for masses of Component C of 45 and 65 $M_\odot$, respectively. We see that in the first case, the ``observable" lifetimes of the simulated trapezia become much shorter than the dynamically determined ones.  Table \ref{tab:OBSERVABLES45} shows that already after one crossing time only 5 trapezia of type (1, 2, 3, 4) and 6 trapezia (1-2, 3, 4) would still be recognizable as such.  Hence, the ``observable lifetime" would be much less than 10,000 years. After 5 crossing times already 61 systems would be  observed as completely dissolved, i.e, consisting only of a tight binary.  Table \ref{tab:OBSERVABLES65} shows results with the larger mass for Component C.  After one crossing time 45 systems would still be observable as trapezia of type (1, 2, 3, 4) while  24 would be trapezia of type (1-2, 3, 4).  After 5 crossing times we have as observable survivors  21 trapezia of type (1, 2, 3, 4) and 20 trapezia of type (1-2, 3, 4).  So, in this case, the``observable lifetimes" would be about 40,000 years.  It is only after 25 crossing times that 50 systems would be observable as completely dissolved, i.e. as consisting only of a close binary.  Again, the ``observable" results for the larger mass of Component C appear more plausible. Figures \ref{fig:OBSERVABLES45} and \ref{fig:OBSERVABLES65} present these ``observable" results.

\begin{table*}
\begin{center}
\caption{Number of ``observable" systems resulting at increasing crossing times for $M_C = 45M_\odot$.}
\begin{tabular}{ccccccc}
\hline

Crossing Times	&	Trapezia (t) 	&   Trapezia (t)   &	Non-Hierarchical	 &	Hierarchical 	&	Hierarchical 	    &	Binaries	\\
                &  (1, 2, 3, 4)     &   (1-2, 3, 4)    &     Triples (NHT)       &   Triples (HT)   &   Quadruples (H)      &      (B)      \\
\hline
                &                   &                  &                         &                  &                       &               \\
          0  	&	 100	        &       0          &	           0	     &	       0	    &        	0	        &     	0	    \\
          1	    &	   5	        &       6          &	          27	     &	      16	    &	        0	        &   	46   	\\
          5   	&	   6            &       3          &	          19         &	      10	    &	        0	        &	    62	    \\
         10	    &	   4	        &       1          &	          17	     &	      10	    &           0	        &	    68   	\\
         25	    &	   1	        &       0          &	           9	     &	       4	    &	        0	        &	    86   	\\
         50	    &	   0	        &       0          &	           1	     &	       7	    &	        0	        &	    92  	\\
        100	    &	   0            &       0          &	           1         &	       2	    &        	0	        &	    97   	\\
\hline

\end{tabular}
\label{tab:OBSERVABLES45}
\end{center}
\end{table*}

\begin{table*}
\begin{center}
\caption{Number of ``observable" systems resulting at increasing crossing times for $M_C = 65M_\odot$.}
\begin{tabular}{ccccccc}
\hline

Crossing Times	&	Trapezia (t) 	&   Trapezia (t)   &	Non-Hierarchical	&	Hierarchical 	&	Hierarchical 	&	Binaries	\\
                &  (1, 2, 3, 4)     &   (1-2, 3, 4)    &     Triples (NHT)      &   Triples (HT)    &   Quadruples (H)  &      (B)      \\
\hline
                &                   &                  &                        &                   &                   &               \\
          0  	&	 100	        &         0        &	      0	            &	    0	        &        	0	    &     	0	    \\
          1	    &	  41	        &        25        &	     19	            &	    7	        &	        0	    &   	8   	\\
          5   	&	  21	        &        20        &	     34            	&	   11	        &	        0	    &	    14	    \\
         10	    &	  13	        &         9        &	     28	            &	   18	        &           0	    &	    32   	\\
         25	    &	   6	        &         2        &	     24	            &	   18	        &	        0	    &	    50   	\\
         50	    &	   2	        &         0        &	     31	            &	    0	        &	        0	    &	    67  	\\
        100	    &	   0         	&         1        &	      4           	&	   15	        &        	0	    &	    80   	\\
\hline

\end{tabular}
\label{tab:OBSERVABLES65}
\end{center}
\end{table*}

\subsection{The Velocity Distribution of the Ejected Stars}
\label{subsec:ejectedstars}

\begin{figure*}
\centering
\includegraphics[width=25.0cm,height=20.0cm]{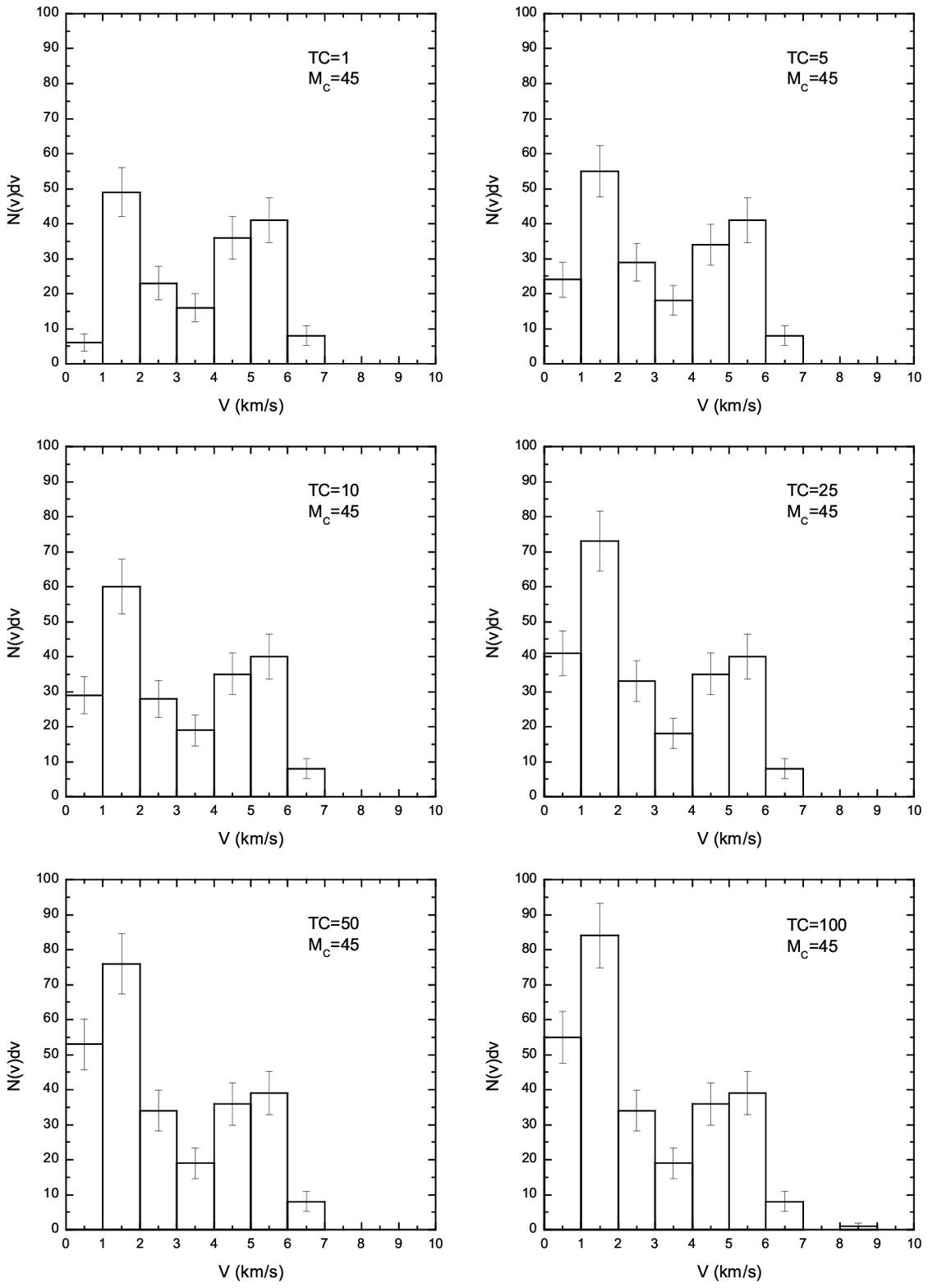}
\caption{Frequency distribution of the velocities of the ejected stars ($M_C = 45M_\odot$). Results are shown after 1, 5, 10, 25, 50 and 100 crossing times. Bars represent statistical uncertainties.}
\label{fig:HISTVELTODASMC45}
\end{figure*}

In the course of the dynamical evolution many stars were ejected from the original trapezia. Figure \ref{fig:HISTVELTODASMC45} displays the distribution of velocities for all ejected stars for $M_C = 45M_\odot$  It is clear from the figure that all ejected stars have relatively low velocities, under 8~km\ s\textsuperscript{-1}. A star was considered to be an ejected escaper when its energy with respect to the center of mass of the system was positive, provided that it did not form a close pair with another star. In all cases considered these escapers left the system permanently.

\begin{figure*}
\centering
\includegraphics[width=10.5cm,height=15.0cm]{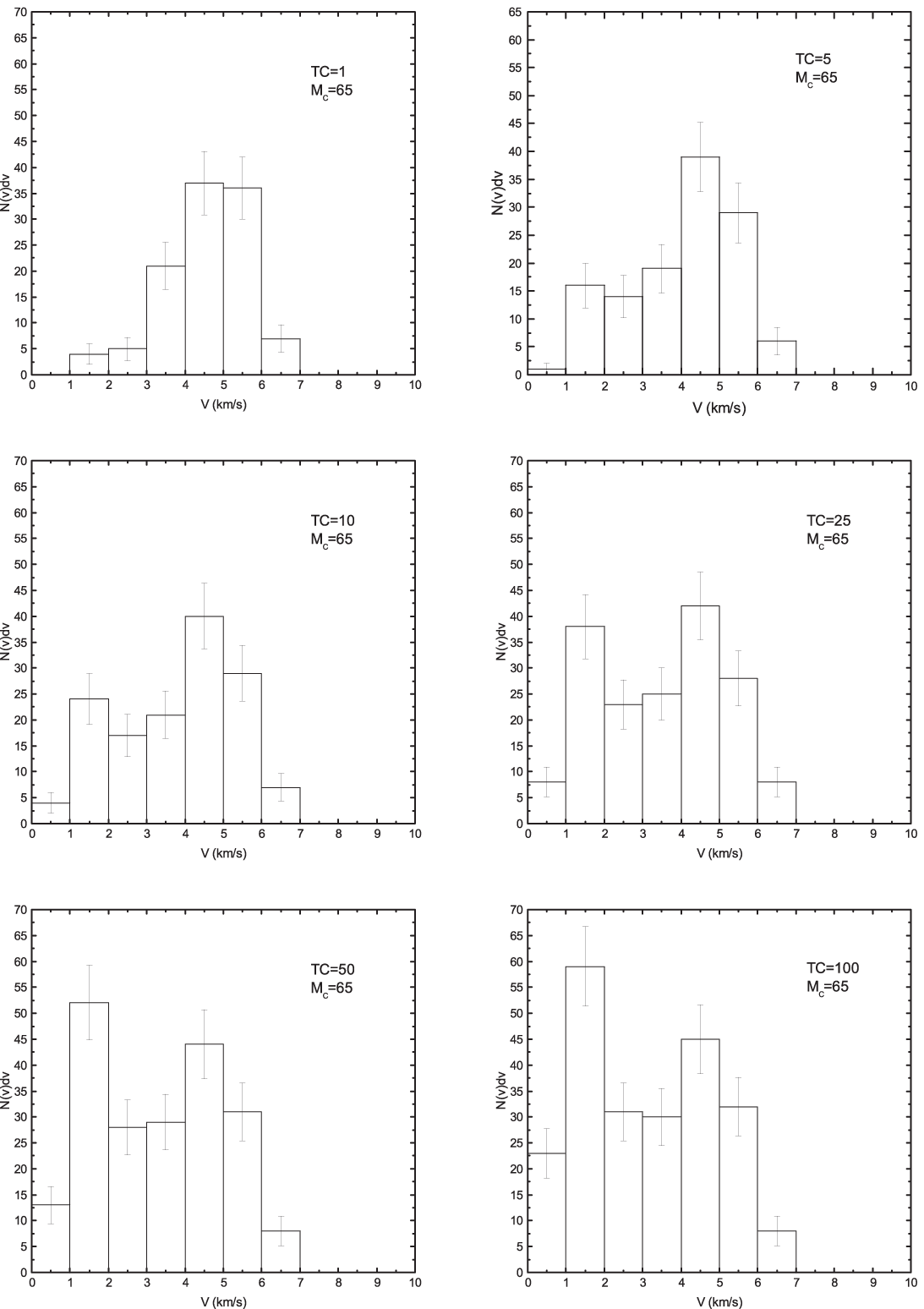}
\caption{Frequency distribution of the velocities of the ejected stars ($M_C = 65M_\odot$). Results are shown after 1, 5, 10, 25, 50 and 100 crossing times. Bars represent statistical uncertainties.}
\label{fig:HISTVELTODASMC65}
\end{figure*}

The velocity distribution of the ejected stars for the ensemble with $M_C = 65M_\odot$ is shown in Figure \ref{fig:HISTVELTODASMC65}.  Again, the escapers have relatively low velocities. It is clear that the encounters occurring during the evolution of the systems are relatively soft, and do not give rise to high velocity escapers, or runaways.  This is in contrast with the situation we found for the minicluster $\theta^1$ Ori B, where close encounters did occur, and whose dynamical evolution did produce a few high velocity escapers.

\subsection{The Properties of the Binaries formed in the Numerical Simulations}
\label{subsec:binariesformed}

\begin{figure*}
\centering
\includegraphics[width=25.0cm,height=20.0cm]{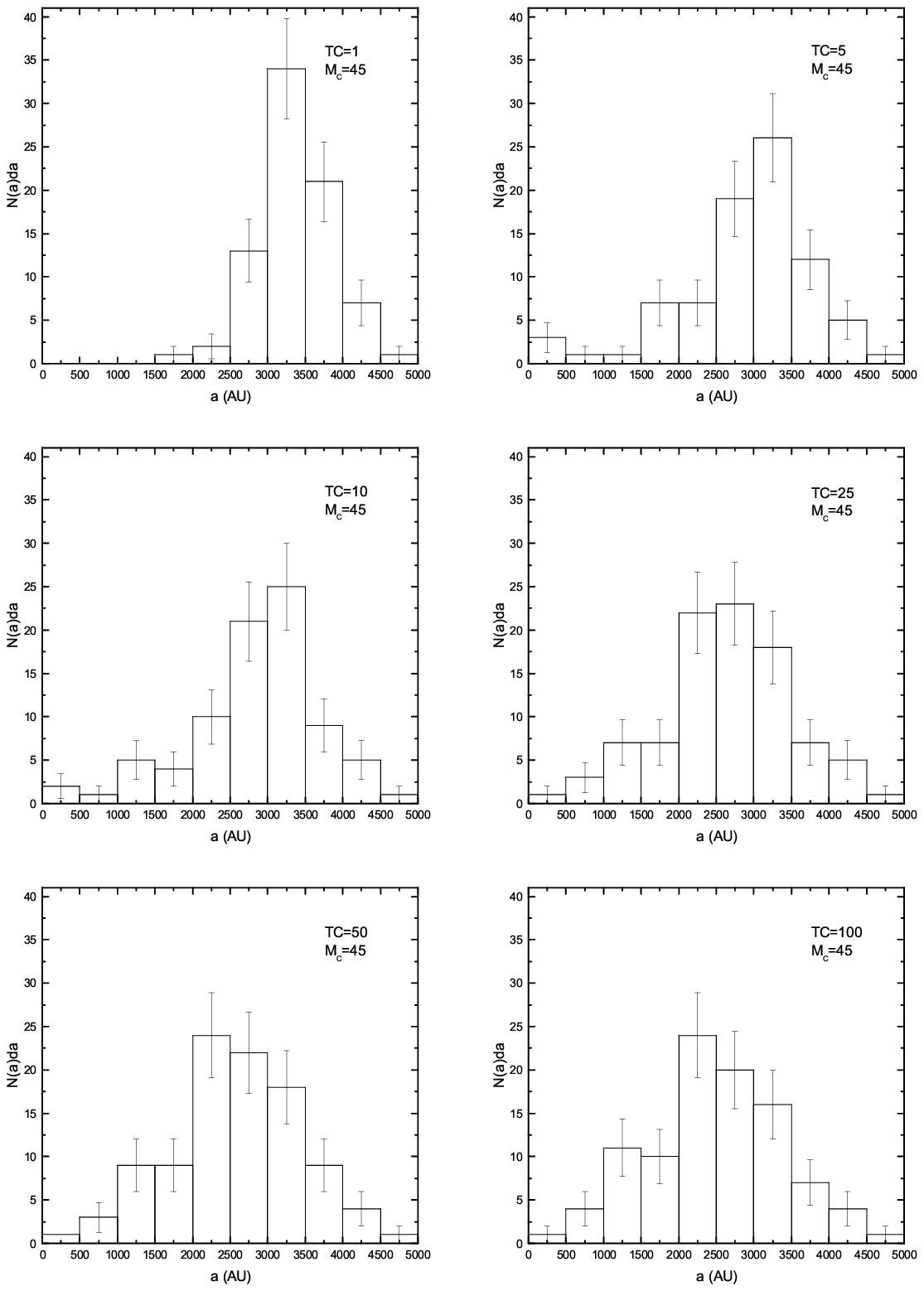}
\caption{Frequency distribution of major semiaxes of the binary systems formed during the integrations ($M_C = 45M_\odot$). Results are shown after 1, 5, 10, 25, 50 and 100 crossing times. Note the sharp maximum at about 2,250 AU. One binary (not shown in the graphs) with $a=10,000$ AU was formed at $CT \lesssim 5 $. It still survived at $CT = 10$ but dissolved before $CT = 25$. }
\label{fig:HISTaTODASMC45}
\end{figure*}

\begin{figure*}
\centering
\includegraphics[width=25.0cm,height=20.0cm]{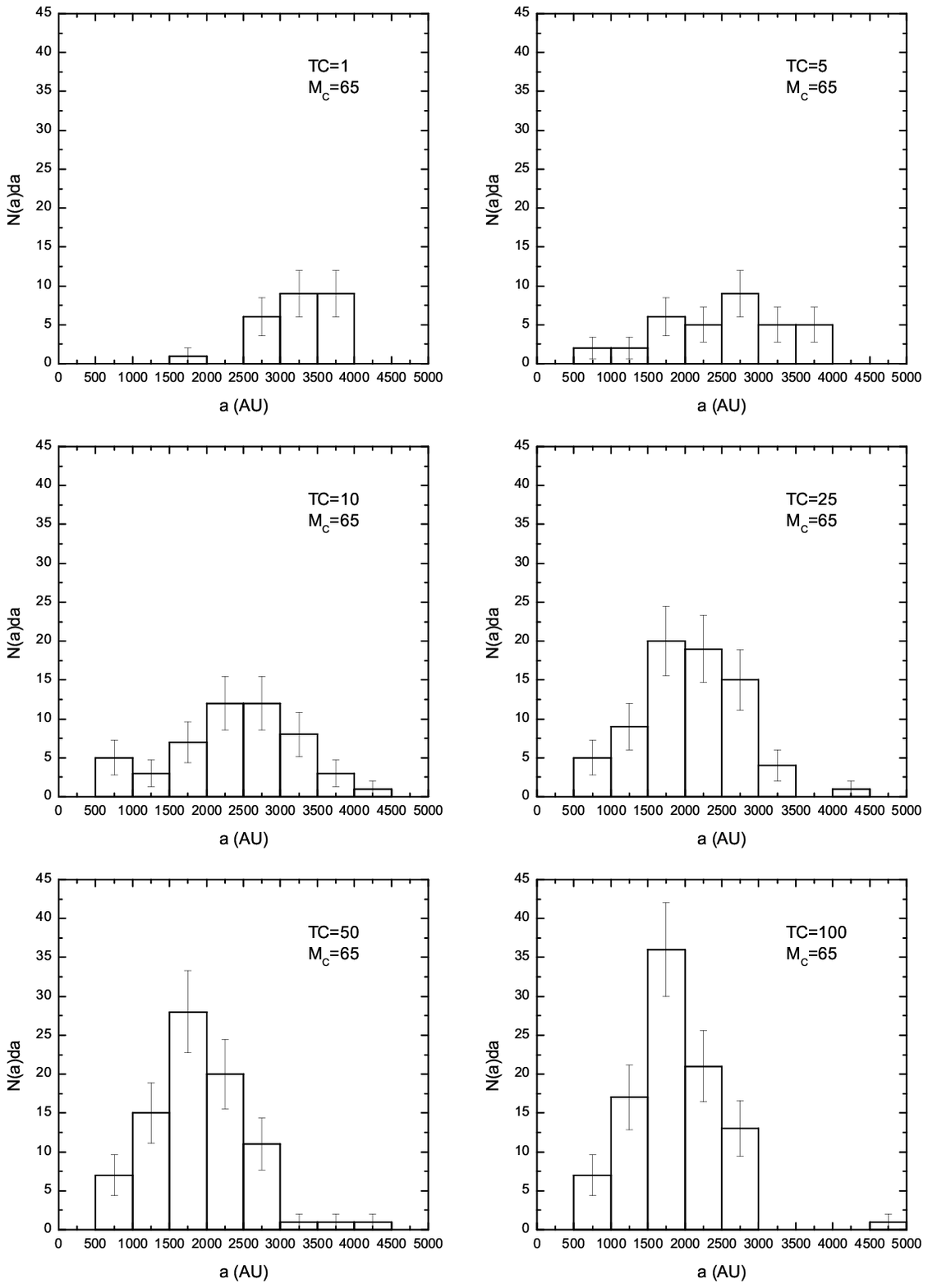}
\caption{Frequency distribution of major semiaxes of the binary systems formed during the integrations ($M_C = 65M_\odot$). Results are shown after 1, 5, 10, 25, 50 and 100 crossing times. Note the sharp maximum at about 1,750 AU.}
\label{fig:HISTaTODASMC65}
\end{figure*}

As the simulated systems reach larger ages, their  dynamical evolution produces an increasing number of bound binaries.  In fact, the most frequent end result of the simulations is a sole binary, which absorbs the binding energy of the initial system, compensating the positive energy carried away by the escapers.  The distribution of major semiaxes of the resulting binaries is shown in Figures \ref{fig:HISTaTODASMC45} and \ref{fig:HISTaTODASMC65}, for values of the mass of Component C of 45 and 65 $M_\odot$, respectively, and for increasing values of the age (in units of the crossing time). The distributions show sharp maxima at $a=2,250$ AU and $a=1,750$ AU respectively. These maxima roughly correspond to the situation where the binding energy of the original trapezium is absorbed by one binary. The dissolution of systems resembling the Orion Trapezium thus produces mostly wide binaries, and may be partly responsible for populating the field with such systems.

\begin{figure*}
\centering
\includegraphics[width=10.5cm,height=15.0cm]{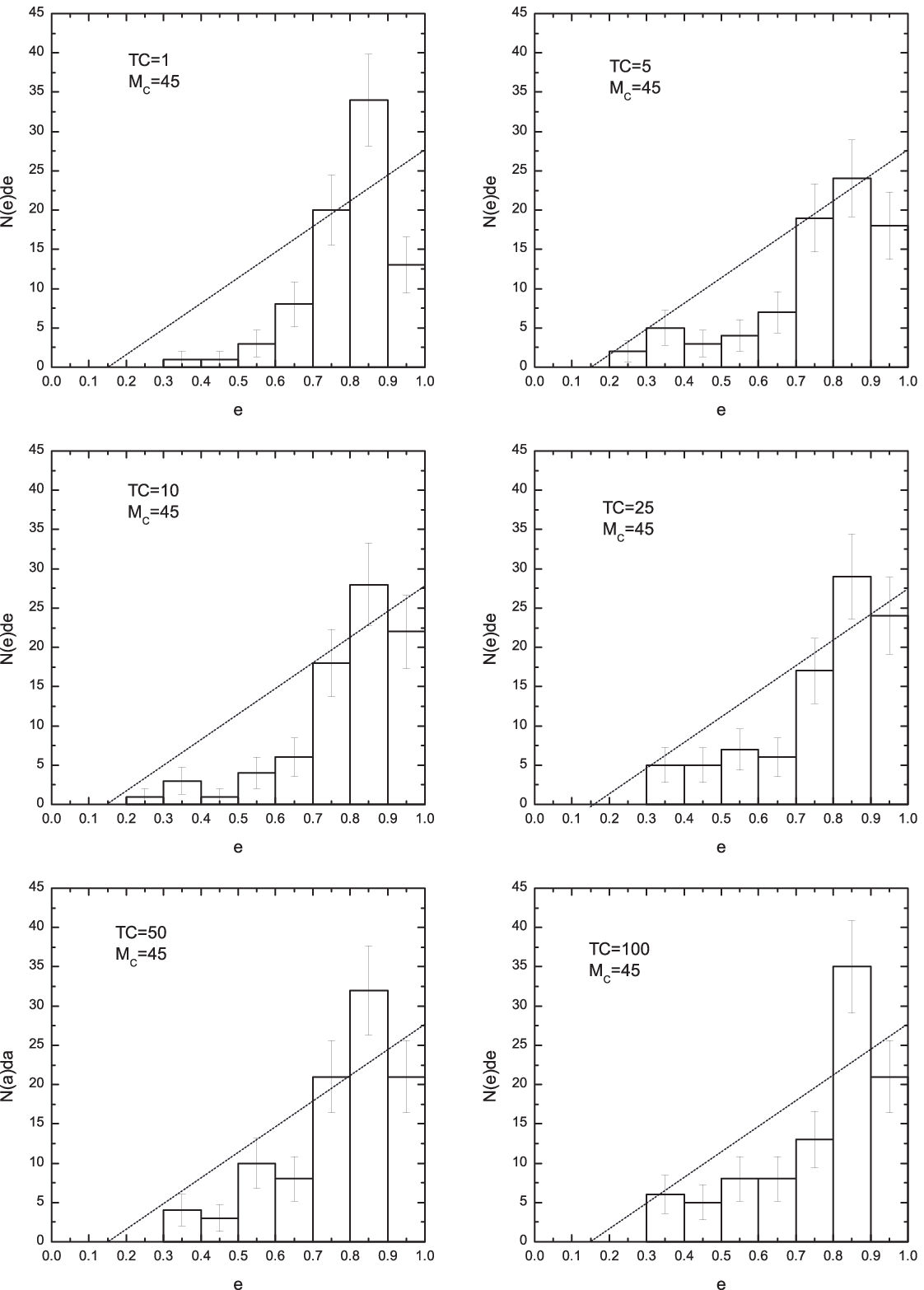}
\caption{Frequency distribution of the eccentricities for the binary systems formed during the integrations ($M_C=45M_\odot$). Results are shown after 1, 5, 10, 25, 50 and 100 crossing times. Note the tendency of the distribution to resemble the thermal one (straight line), i.e. $N(e)$d$e=ke$d$e$.}
\label{fig:HISTeTODASMC45}
\end{figure*}

\begin{figure*}
\centering
\includegraphics[width=10.5cm,height=15.0cm]{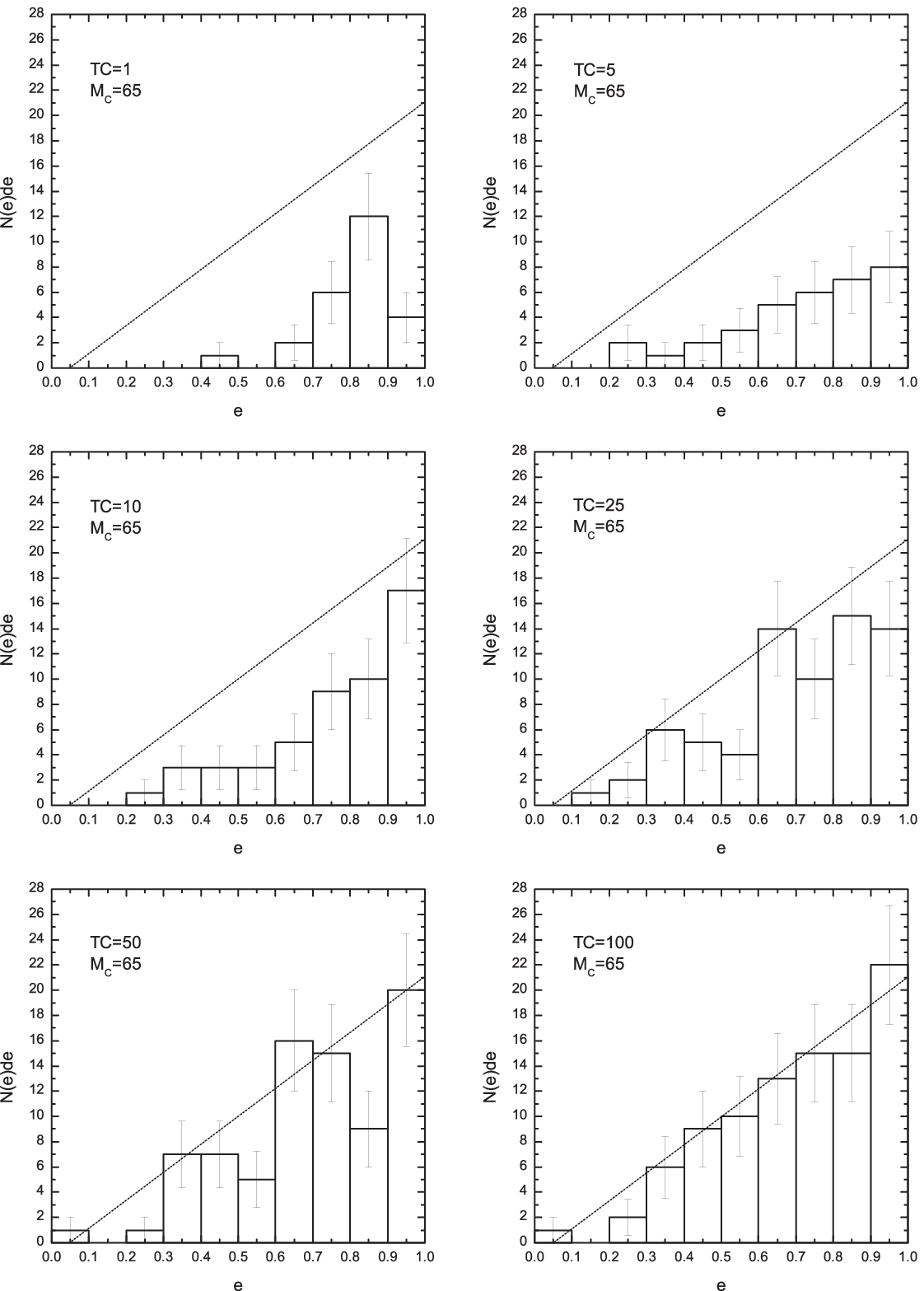}
\caption{Frequency distribution of the eccentricities for the binary systems formed during the integrations ($M_C=45M_\odot$). Results are shown after 1, 5, 10, 25, 50 and 100 crossing times. Note the tendency of the distribution to resemble the thermal one (straight line), i.e. $N(e)$d$e=ke$d$e$.}
\label{fig:HISTeTODASMC65}
\end{figure*}

The distribution of eccentricities is shown in Figures \ref{fig:HISTeTODASMC45} and \ref{fig:HISTeTODASMC65}, again for values of the mass of Component C of 45 and 65 $M_\odot$, respectively, and for increasing values of the age (in units of the crossing time). The average value of the eccentricities after 100 crossing times is equal to 0.77 for the cases with $M_C = 45M\odot$, and 0.71 for those with $M_C = 65M\odot$, close to, but slightly larger than, the expected value for a group of binaries whose distribution in phase space depends only on their energy (Ambartsumian, 1937).  Figures~\ref{fig:HISTeTODASMC45} and \ref{fig:HISTeTODASMC65} also show that the eccentricity distribution follows approximately the relation $f(e)$d$e = ke$d$e$ (Heggie, 1975), that is, it is roughly thermal.

\begin{figure}
\centering
\includegraphics[width=10.0cm,height=7.0cm]{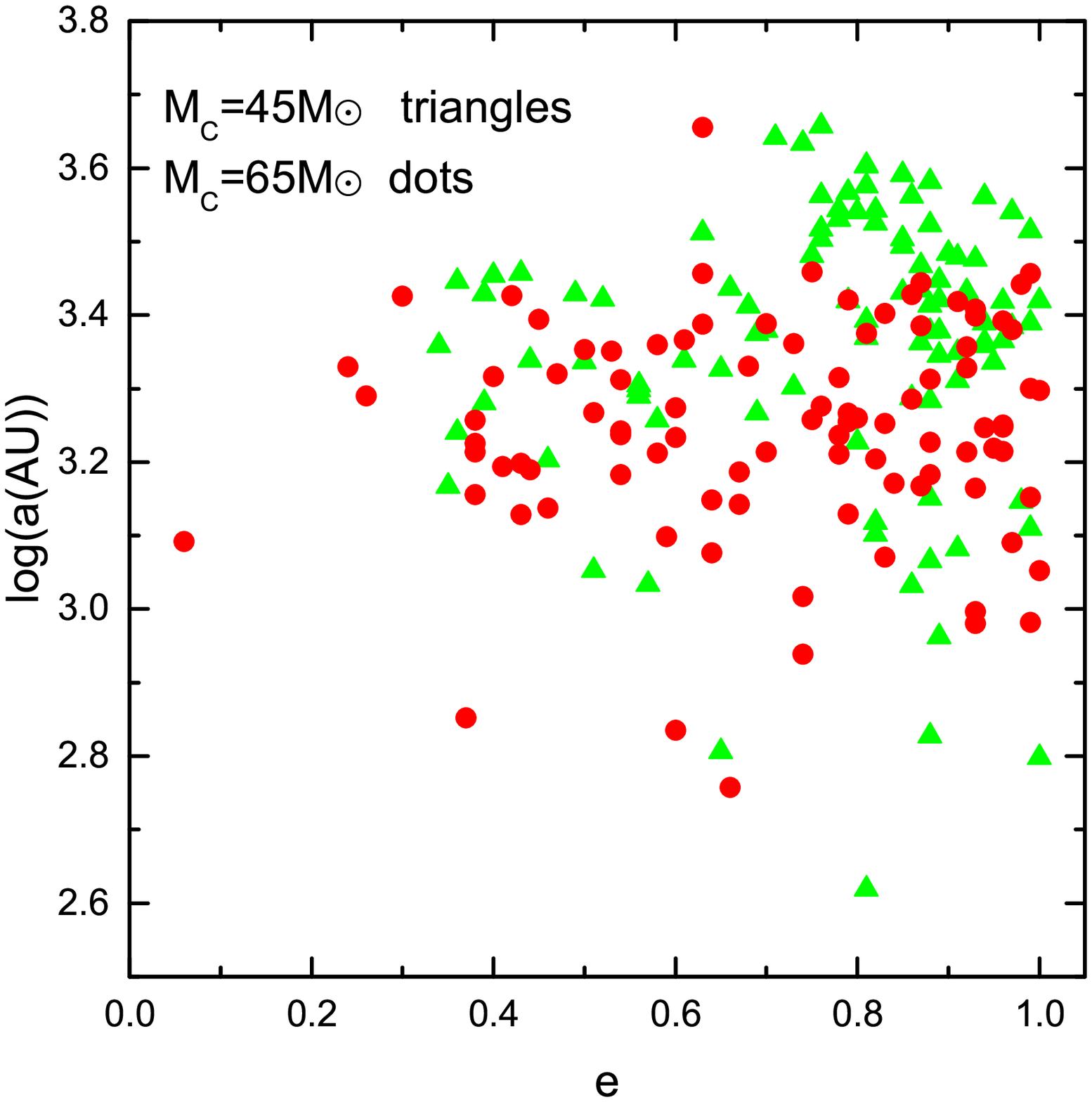}
\caption{Major semiaxes as a function of eccentricities for all the binaries formed in the numerical integrations. Triangles correspond to the case with $M_C = 45M_\odot$ and dots to the case with $M_C = 65M_\odot$. Very few binaries with eccentricities smaller than 0.3 are present, an expected result for binaries formed via dynamical interactions. Large eccentricities are present predominantly at large values of the semiaxes. The majority of semiaxes cluster around 2,000 AU.}
\label{fig:LOGAvsE}
\end{figure}

Figure~\ref{fig:LOGAvsE} displays the major semiaxes as a function of the eccentricities for a total of $98$ and $95$ binaries formed during the integrations at $CT = 100$ for both values of $M_C$. This figure resembles the observational result of Raghavan et al. (2010) for the region of long period $(P \geq 10^4$ days). Very few binaries appear to have eccentricities $e \leq 0.3$ at all values of the major semiaxis. For larger values of $e$, the points seem to be randomly scattered in the vertical direction of the plot but still in a limited region of all possible binary major semiaxes.

\section{Discussion and Conclusions}
\label{sec:conclusions}

Our numerical exploration of ensembles of systems mimicking the dynamical evolution of the Orion Trapezium has produced some remarkable results.  We can summarize them as follows.

\begin{itemize}
	\item Star E nearly always escapes right at the beginning of the numerical integrations.  In 300 runs conducted, we found only one exception, where Star E became bound for a short period but escaped soon afterwards.
 	\item Including Star E all systems turned out to have  a small positive virial value. Excluding Star E, the virial value was approximately zero.
	\item  In the time-reversal runs, Star E was captured in about 25 \% of the cases in less than 2,000 years.  This suggests that Star E is probably a recently ejected member of the Orion Trapezium.
	\item Using the best available value for the mass of Component C, only 25\% of the simulated  systems survived as trapezia after 10,000 years, but only 5\% resembled the original trapezium configuration.  Therefore, the mean lifetime would be much less than 10,000 years, a value that seems implausibly short for the Orion Trapezium.
	\item With a larger mass for the C component the mean dynamical lifetime of the systems turned out to be between 10,000 and 50,000 years.  This value is compatible with the dynamical age we estimated for the minicluster $\theta^1$ Ori B (about 30,000 years).  It is also compatible with the age of the ``Huygens region" of the Orion Nebula, estimated to be 15,000 years (O'Dell et al. 2009). We think therefore that the value of 45 $M_\odot$ for the mass of Component C is an underestimate.
	\item The end result of the dynamical evolution after 100 crossing times was usually a tight binary, sometimes a hierarchical triple system.
	\item The properties of the binaries formed during the integrations are comparable with observational values.  The eccentricity distribution of the binaries formed in the numerical simulations is approximately thermal. The dissolution of systems closely mimicking the Orion Trapezium produces mostly wide binaries and could be partly responsible for populating the field with such massive systems. The distribution of the major semiaxes has a maximum roughly corresponding to the binding energy of the initial systems.
    \item The period-eccentricity distribution closely resembles the observational data. Very few binaries with $e \leq 0.3$ are produced for all values of $a$.
	\item Most of the ejected stars have velocities close to the escape velocity.  No runaway stars were produced in the simulations.  This means that it is highly unlikely for systems resembling the Orion Trapezium as a whole to be able to generate runaway stars, as has sometimes been claimed.  We recall that we did find some runaways in our previous study of the dynamical future of the minicluster $\theta^1$ Ori B (Allen et al., 2015).  The evolution of this mini-cluster is obviously more violent than that of an entire trapezium.
\end{itemize}

A more complete numerical exploration of the possible outcomes of the dynamical evolution of systems mimicking the Orion Trapezium should include a much larger number of numerical realizations.  However, we believe that we have presented a fair sampling of possible initial conditions obtained from the best currently available observational data. We disregarded the PA rates of change because of their low accuracy. This implies that we are underestimating the tangential velocities, which in turn means that the dynamical lifetimes should be somewhat shorter. We tried to be as explicit as possible about the uncertainties in the observational data and about the assumptions implicit in their derivation. It is probably not worthwhile to conduct numerical experiments with more realizations until more reliable values for the tangential and radial velocities as well as for the masses of the components become available.

 An interesting question is whether the widely accepted notion that the Orion Trapezium is a bound system is valid, especially in view of the rather short lifetimes we find, and of the escaping Component E. At least two arguments support its being a bound system. First, the probability of finding four bright stars within a radius of $10$ arcseconds is very low. Second, the relative motions of the main components in the plane of the sky are very small, also implying a bound system. However, if a better knowledge of the radial velocities should show widely discrepant values, the question would have to be re-examined.

\vskip3.0cm

\section*{Acknowledgments}

We thank the support provided by Instituto de Astronom\'{\i}a at Universidad Nacional Aut\'onoma de M\'exico (UNAM).
We would like to thank Juan Carlos Yustis for his help with the figures presented in this paper. We also thank Direcci\'on General de Asuntos del Personal Acad\'emico, DGAPA at UNAM for financial support under projects number PAPIIT IN103813, IN102617 and IN102517. The constructive remarks of an anonymous referee are gratefully acknowledged.

\end{document}